\documentclass[11pt]{article}
\usepackage{fa2025}
\usepackage{amsmath}
\usepackage{url}
\usepackage{graphicx}
\usepackage{color}
\usepackage{soul}
\usepackage{siunitx}
\usepackage{subcaption}
\usepackage[utf8]{inputenc}
\usepackage[backend=biber, maxbibnames=3, minbibnames=1]{biblatex} % maxbibnames controla la cantidad en la bibliografía
\AtEveryBibitem{%
  \clearfield{url}% Elimina el URL
  \clearfield{urldate}% Elimina la fecha de visita
  \clearfield{doi}
}
\renewbibmacro*{url+urldate}{%
  \printfield{url}} % Solo imprime el campo URL
\AtEveryBibitem{\clearname{editor}}

\AtEveryBibitem{\clearfield{note}}

\title{Open-Source System for Multilingual Translation and Cloned Speech Synthesis}

\multauthor
{Mateo Cámara$^{1,2*}$ \hspace{1cm} Juan Gutiérrez$^1$ \hspace{1cm} María Pilar Daza$^1$ \hspace{1cm} José Luis Blanco$^{1,2}$} { \\
  $^1$ETSI de Telecomunicación, Universidad Politécnica de Madrid, Spain\\
$^2$Information Processing and Telecommunications Center, Universidad Politécnica de Madrid, Spain
\correspondingauthor{mateo.camara@upm.es}{Mateo Cámara et al.}
\thanks{
M. Cámara, J. Gutiérrez, M.P. Daza, and J.L. Blanco, are with the Signal Processing Applications group (GAPS-UPM). 
These activities were partially funded by the Horizon 2020 Research and Innovation EU Programme under grant agreement No. 101003750; and by the Ministry of Economy and Competitiveness of Spain under grant No. PID2021-128469OB-I00. 
}
}

\begin{document}
\defbibheading{mybib}{%
  \section{References}% Sección sin número
  \addcontentsline{toc}{section}{References}% Opcional: añade al índice
}

\maketitle
\begin{abstract}
We present an open-source system designed for multilingual translation and speech regeneration, addressing challenges in communication and accessibility across diverse linguistic contexts. The system integrates Whisper for speech recognition with Voice Activity Detection (VAD) to identify speaking intervals, followed by a pipeline of Large Language Models (LLMs). For multilingual applications, the first LLM segments speech into coherent, complete sentences, which a second LLM then translates. For speech regeneration, the system uses a text-to-speech (TTS) module with voice cloning capabilities to replicate the original speaker’s voice, maintaining naturalness and speaker identity.

The system’s open-source components can operate locally or via APIs, offering cost-effective deployment across various use cases. These include real-time multilingual translation in Zoom sessions, speech regeneration for public broadcasts, and Bluetooth-enabled multilingual playback through personal devices. By preserving the speaker’s voice, the system ensures a seamless and immersive experience, whether translating or regenerating speech.

This open-source project is shared with the community to foster innovation and accessibility. We provide a detailed system performance analysis, including latency and word accuracy, demonstrating its potential to enable inclusive, adaptable communication solutions in real-world multilingual scenarios.
\end{abstract}

\section{Introduction}\label{sec:introduction}

In an era where global collaboration and mobility are redefining human interaction, the ability to communicate across languages in real time has transitioned from a futuristic vision to a technological imperative. Traditional translation pipelines, reliant on sequential human intervention or disjointed automated tools, are ill-suited for dynamic environments—from international conferences to emergency response scenarios—where managing latency and coping with naturalness is critical \cite{yan_cmus_2023}. Artificial intelligence (AI)-powered speech translation systems promise to address this gap. Yet, their practical use remains constrained by competing priorities: the need for accuracy, computational efficiency, and the preservation of speaker identity \cite{iranzo-sanchez_streaming_2021, barrault_joint_2025}. While recent breakthroughs in neural architectures have accelerated progress, the field remains divided between two competing paradigms—end-to-end and cascade systems—each with distinct trade-offs in performance, adaptability, and scalability \cite{agarwal_findings_2023}.  

End-to-end speech translation systems, which directly map source-language speech to target-language text or audio, have gained prominence for their potential to minimize error propagation through joint optimization of transcription and translation tasks. Pioneering works \cite{berard_end--end_2018, jia_direct_2019, lavie_janus-iii_1997, lee_direct_2022} demonstrated the feasibility of unified frameworks implemented on neural networks, while recent breakthroughs like Meta’s SeamlessM4T \cite{barrault_joint_2025} have elevated the paradigm by supporting multilingual translation across 100+ languages. SeamlessM4T exemplifies the allure of end-to-end architectures: it eliminates modular bottlenecks by training on massive multimodal corpora, achieving state-of-the-art BLEU scores \cite{papineni_bleu_2002} while preserving prosody. However, as Bahar et al. \cite{bahar_comparative_2019} note, such systems remain constrained in streaming applications due to their reliance on full input sequences. The same ability to cope with extended semantic contexts, from which they excel in performance, makes them prone to latency spikes in real-time scenarios. Moreover, end-to-end models struggle with interpretability and robustness, mainly when dealing with noisy input or domain shifts. Anyhow, SeamlessM4T is not yet publicly available, and its real-world applicability beyond controlled demonstrations remains to be thoroughly evaluated.

By contrast, cascade systems modularize translation into discrete stages—ASR, text refinement, Machine Translation (MT), and TTS—leveraging specialized models optimized for each subtask. This decoupled architecture, as explored by Iranzo-Sánchez et al. \cite{iranzo-sanchez_streaming_2021} and Arivazhagan et al. \cite{arivazhagan_monotonic_2019}, benefits from abundant monolingual ASR and MT datasets, enabling robust performance in high-resource settings. However, cascade pipelines inherit vulnerabilities from the accumulation of errors between stages and segmentation challenges in continuous speech streams \cite{fungeneo, oda2014optimizing}. Despite these challenges, recent evaluations confirm that cascade systems retain a performance edge over end-to-end models in standardized benchmarks, particularly for high-resource language pairs \cite{agarwal_findings_2023}. Like those by Bahar et al. \cite{bahar_start-before-end_2020}, hybrid solutions integrate streaming ASR with simultaneous MT models to mitigate latency. Their reliance on proprietary components limits reproducibility, a critical barrier for academic and low-budget implementations.

\begin{figure*}[h]
    \centering
    \includegraphics[width=1\linewidth]{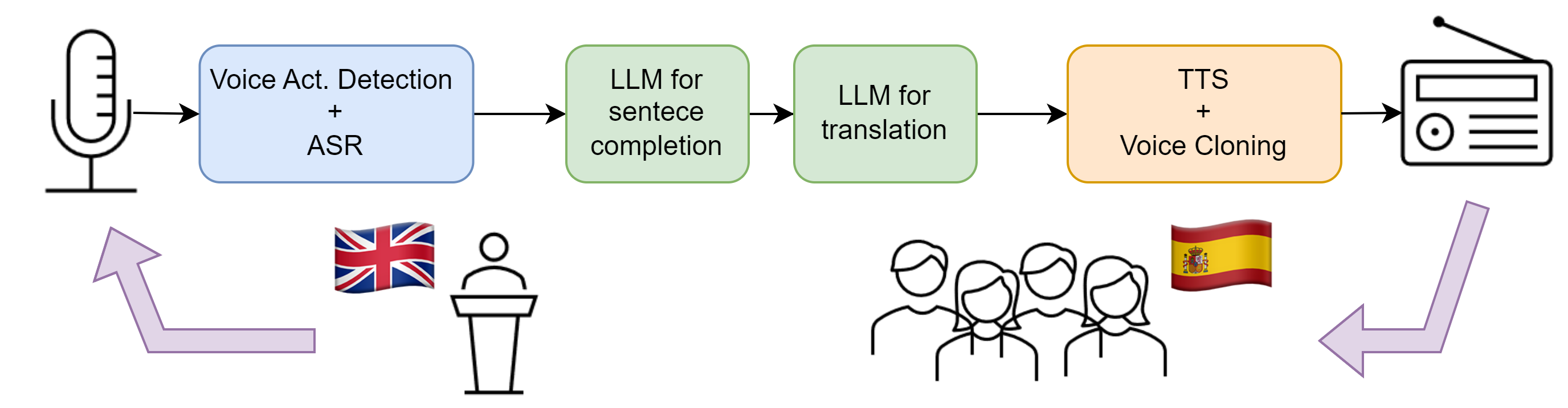}
    \caption{Schema of the solution for multilingual speech recognition (left), translation, and synthesis (right).}
    \label{fig:main-diag}
\end{figure*}

This work presents a fully open-source, modular cascade system designed for conferences to harmonize the strengths of pre-trained models while mitigating their weaknesses and delivering a complete end-to-end experience. Figure~\ref{fig:main-diag} shows the schematic of our solution. Implements a four-stage pipeline. Unlike end-to-end frameworks, our system emphasizes transparency and adaptability by integrating best-performant, independently validated, state-of-the-art open-source modules. We selected the following for our implementation: 1) Whisper \cite{radford_robust_2023} for low-latency ASR; 2) a Llama 3 model \cite{grattafiori_llama_2024} serves as a contextual buffer, correcting or substituting misrecognized phonemes and deciding whether the sentence is ready for translation, by leveraging contextual and syntactic cues; 3) a second Llama 3 LLM conditioned to perform translation; and 4) MeloTTS \cite{zhao2024melo}, offline fine-tuned for voice cloning to mimic the speaker. 

We have developed and publicly released this solution for general use,\footnote{\url{https://github.com/MateoCamara/speech-translator-with-voice-cloning}} including detailed descriptions of the system and demo's software and hardware implementations. Specifically, we created a low-power analog radio transmitter setup to take advantage of the limits of commercial FM band frequencies, allowing easy reception through standard consumer radio devices. The design simplifies accessibility, accommodating users less comfortable with advanced technology. Additionally, we provide clear instructions for integrating the system with video conferencing sessions and Bluetooth-enabled devices, enhancing its versatility and user-friendliness.

We have validated this solution objectively by measuring latency (between both ends), Word Error Rate (WER) for ASR, and BLEU \cite{papineni_bleu_2002} and COMET \cite{rei_comet_2020} scores for translation quality. Regarding the speech synthesis module, we conducted a subjective evaluation of the voice cloning component using a speaker and a small group of evaluators. While this subjective evaluation is informal and not exhaustive in terms of languages or vocabulary, we deem it sufficient for our demonstration.

The remainder of this document is as follows. Section \ref{sec:software} describes the software development aspects, including system architecture and implementation details. Section \ref{sec:hardware} covers the hardware development, particularly the analog radio setup designed for ease of use. Section \ref{sec:experiments} discusses the tests conducted and the validation metrics obtained to ensure the system's reliability and effectiveness. Finally, Section \ref{sec:conclusion} concludes the document.

\section{Software Implementation} \label{sec:software}

\subsection{Automatic Speech Recognition (ASR)}

The ASR module forms the foundational layer of our cascade system, delivering timely transcriptions from the input speech. For this module, we use Whisper \cite{radford_robust_2023}, an open-source, multilingual model pre-trained on five million hours of diverse labeled audio data. We selected it for its proven robustness to accents, background noise, and streaming compatibility. Unlike proprietary alternatives (e.g., Google Speech-to-Text, Amazon Transcribe), Whisper’s architecture and weights are publicly accessible. Specifically, we utilize whisper.large-v3.turbo\footnote{\url{openai/whisper-large-v3-turbo}} (1,550M parameters), which achieves a WER of 9.5\% on average for the 15 most common languages, just 1\% higher than the non-distilled version of the model, being 5.4 faster on inference. 

A critical limitation of conventional ASR systems in real-time applications is in managing non-speech intervals (e.g., silence, background noise), wasting computational resources, and introducing additional latency with spurious outputs. To minimize computational overhead and false activations, we integrated Silero VAD \cite{Silero_vad}, a lightweight, open-source voice activity detector optimized for real-time applications. Silero VAD processes audio 30 times faster than real-time on CPU, achieving an accuracy score of 91\% on their validation dataset. We use it to activate Whisper only when speech probability exceeds a threshold of 0.5, reducing idle inference cycles. %This approach avoids processing non-speech intervals, a critical optimization for resource-constrained deployments.

\subsection{LLMs for Context Refinement and Translation}

To address the inherent challenges of real-time speech processing—such as disfluencies, fragmented phrases, and contextual ambiguity—we deploy two specialized LLMs in cascade. This modular design ensures robustness while balancing latency and semantic coherence.

\subsubsection{Context-Aware Phrase Refinement}

The first LLM acts as a linguistic validator tasked with resolving ambiguities, correcting ASR errors, and determining phrase completeness. Built on LLama-3 architecture \cite{touvron_llama_2023}, the module leverages a five-sentence context stored in a buffer for semantic continuity. We found this effectively complements Whisper's 30-second speech context. Specifically, we use the LLaMA 3.3-70B-Instruct\footnote{\url{meta-llama/Llama-3.3-70B-Instruct}} model, which offers an optimal balance between quality and inference speed, achieving 51.14 tokens per second. As a new chunk of transcription arrives from Whisper, this LLM is prepared to account for:

\begin{itemize}
    \item Completeness: Whether the chunk forms a grammatically and semantically self-contained unit (e.g., ending with punctuation or a natural pause).
    \item Anomaly Detection: Identifies and removes extraneous elements (e.g., filler words, misrecognized phonemes) using syntactic and contextual cues.
\end{itemize}

The module retains incomplete phrases in a rolling buffer with a capacity of five chunks. If the buffer reaches its maximum capacity, the chunks are forcibly flushed to the translation module, prioritizing low latency over prosodic perfection.

\subsubsection{Translation}

The second LLM handles translation, converting validated phrases into the target language. Built on LLaMA 3.3-70B, it offers native high multilingual performance, supporting eight languages, including English, Spanish, French, and German. Each complete phrase from the previous LLM is translated, preserving semantic integrity. 

Alternative implementations may merge these two LLM instances. For the sake of simplicity and at the cost of increased latency, we did not address such considerations, leaving them for future work. 

\subsection{Text-to-Speech with voice cloning}

To bridge the gap between translated text and natural-sounding speech, our system integrates a hybrid TTS pipeline that combines high-quality synthesis and voice cloning for a specific speaker. This dual-stage approach ensures linguistic clarity and preserves the original speaker’s vocal identity. 

\subsubsection{MeloTTS for Speech Synthesis}

For the text-to-speech process, we employ MeloTTS \cite{zhao2024melo}, an open-source, non-autoregressive TTS framework optimized for low-latency streaming. It supports real-time inference (i.e., speech synthesis) on both CPU and GPU, delivering 44.1 kHz audio with high fidelity. Despite its quality, the model remains efficient, operating smoothly even on mid-range hardware. MeloTTS leverages non-autoregressive architectures, allowing for single-step audio generation. It supports multiple languages and accents, including English, Spanish, French, Chinese, Japanese, and Korean. Additionally, it enables speech rate adjustment, offering flexibility for various applications.

\subsubsection{Voice Cloning Via Full Retraining}

Unlike methods that adapt pre-trained models through style transfer (e.g.,  ``voice color'' tuning with 30 seconds of audio), we adopt a full retraining strategy to clone speaker identities with high fidelity. This approach addresses the limitations of zero-shot cloning systems, which often struggle to preserve personal vocal traits (e.g., breathiness, pitch contours) in low-resource scenarios. The implementation workflow is as follows:

\begin{itemize}
 \item Base Model Selection: We initialize MeloTTS with weights pre-trained on a target-language corpus (in the case of Spanish, 25 hours of clean and consistent audio), prioritizing phonetic coverage of the target deployment language.
 
 \item Speaker-Specific Retraining: We used 30 minutes of clean audio from the target speaker (industry-standard duration) to fine-tune the model parameters. For the MeloTTS, these included the audio generator, the duration predictor, and the audio discriminator models. Based on our tests, the latter could be frozen, accelerating the process. 
\end{itemize}

% Training during 2 days and 12 hours
% IPTC gpu - 
We trained our base model for a male Spanish speaker on the LibriVox datasets \cite{kearns_librivox_2014}. The retraining of the speaker-specific model was covered on a NVIDIA A100@40GB GPU for 56 hours. The process stopped after 235k epochs as the losses stabilized.

\section{Hardware Implementation}\label{sec:hardware}

To ensure inclusivity across diverse user demographics—particularly those with limited technological literacy, such as elderly populations—our system prioritizes analog FM radio transmission while retaining compatibility with modern digital interfaces. The FM-based design guarantees universal accessibility through low-cost, widely available receivers (e.g., portable transistor radios). Bluetooth multicast and virtual audio routing remain as supplementary options for tech-literate users.

\subsection{Analog FM Radio for Universal Deployment}

The core hardware deployment lies in a compact, low-power FM transmitter designed to operate at the edges of commercial frequency bands (87.5-108 MHz). It allows reception on conventional devices while avoiding interference with licensed broadcasts. The system comprises three key components:

\begin{itemize}
    \item Transmitter Module: We included a DSP PLL FM transmitter module operating at carrier frequency 105.1~MHz with a default output power of 0.5~W.
    \item Amplification Stage: A Class-A amplifier (gain ×100, 12 V input) boosts the signal to 2.0~W, enabling coverage of large indoor spaces (e.g., auditoriums).
    \item Antenna System: A vertically polarized mini patch panel antenna with +3~dB gain, 50~$\Omega$ impedance ensures robust coverage while minimizing signal leakage beyond 100 m.
\end{itemize}

\begin{figure}
    \centering
    \begin{subfigure}{0.49\linewidth}
        \centering
        \includegraphics[width=\linewidth]{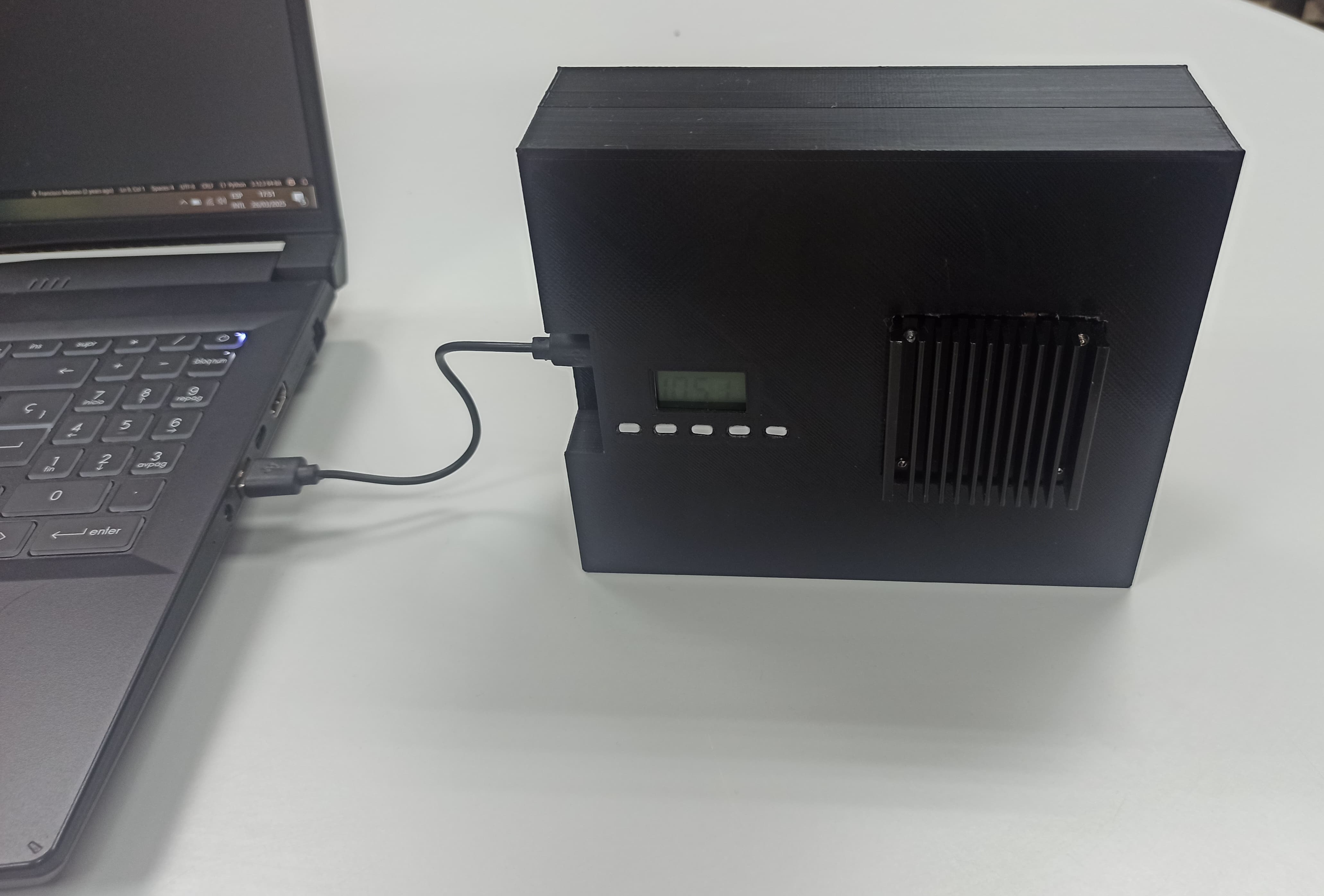}
    \end{subfigure}
    \hfill
    \begin{subfigure}{0.49\linewidth}
        \centering
        \includegraphics[width=\linewidth]{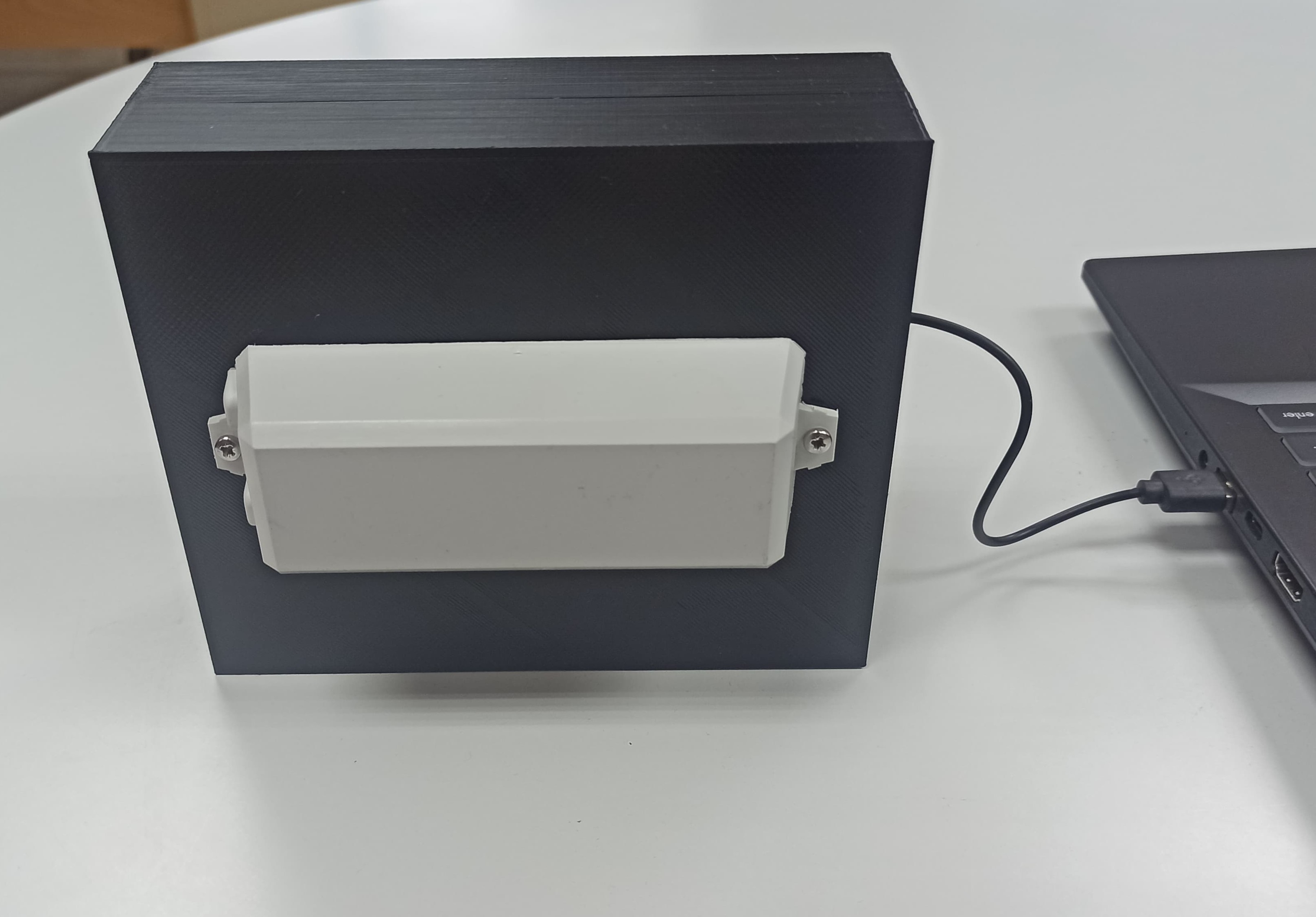}
    \end{subfigure}
    \caption{Radio Transmitter and laptop.}
    \label{fig:hardware-display}
\end{figure}

% The transmitter achieved a 50 m indoor range in our tests, sufficient for most public venues. The integrity of the radio signal was verified using an RTL-SDR dongle, showing a carrier-to-noise ratio (CNR) of \hl{XXXXXXXXXX} dB at a distance of 20 m, adequate for intelligible audio. 
It supports 12 V DC input from batteries or wall adapters. Thermal fuses and current limiters prevent overheating. Figure~\ref{fig:hardware-display} shows a picture of the hardware display used in our tests, housed in a PBS enclosure. The system offers versatile audio input via a 3.5 mm jack, micro-USB, or built-in microphone. It features an LCD screen that displays transmission information with volume, play/pause, and frequency selector buttons.

\subsection{Complementary Digital Interfaces}

For users with smartphones or computers, the system optionally routes translated audio to Bluetooth headphones (via multicast pairing) or virtual microphone outputs for video conferencing tools. However, these interfaces are secondary to the FM radio, which remains the primary channel for its simplicity, cost-effectiveness, and compatibility with legacy devices. The associated GitHub page provides an explanation of how to set these up.

\section{Experiments and Evaluation}\label{sec:experiments}

We evaluated our solution using the Europarl dataset \cite{iranzo-sanchez_europarl-st_2020}, which contains recordings from the European Parliament. Speeches are delivered and translated into multiple languages. We selected the test subset for validation and assessed the system’s performance in translating from English to Spanish on up to three hours of speech and the corresponding transcripted sentences. 

The analysis included hereafter covers several key aspects: system latency, errors in speech recognition and translation, voice cloning quality (for a single user), and the performance of the hardware radio setup. 

\subsection{Latency Tests}

\begin{figure}
    \centering
    \includegraphics[width=1\linewidth]{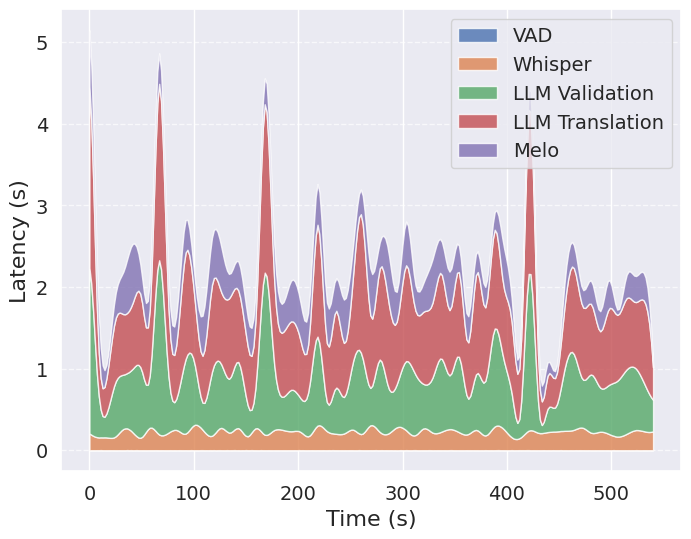}
    \caption{Results on the latency test with 10 minutes speech including all pipeline modules.}
    \label{fig:latency}
\end{figure}

This experiment quantifies the latency from the moment a speaker begins a sentence until playback starts. Figure~\ref{fig:latency} presents an example of a 10-minute speech, illustrating latency variations over time. This fragment is the first eight audio files of the Europarl test set, consisting of political speeches delivered by different speakers. These speeches offer a variety of vocal characteristics, accents, and speaking styles. The total latency averages around 2.5 seconds, with peaks reaching 5 seconds, highlighting the system's behavior with diverse input signals. Most components remain within a stable range. The literature considers 2.5 seconds a valid threshold for real-time translation \cite{agarwal_findings_2023}.

Whisper and MeloTTS ran in parallel on an RTX 5090 GPU, while LLM tasks relied on cloud-hosted A100 units accessed via API. The primary latency contributors are the LLM-related tasks, which are challenging to control due to shared cloud resources. Dedicated inference services could help reduce variability. The peak latencies are infrequent and not representative of typical system performance. Latency would likely increase significantly on less powerful hardware. Therefore, deploying cloud-based resources similar to those used in our experiments is recommended.

\subsection{Intelligibility tests}

\begin{figure}
    \centering
    \includegraphics[width=1\linewidth]{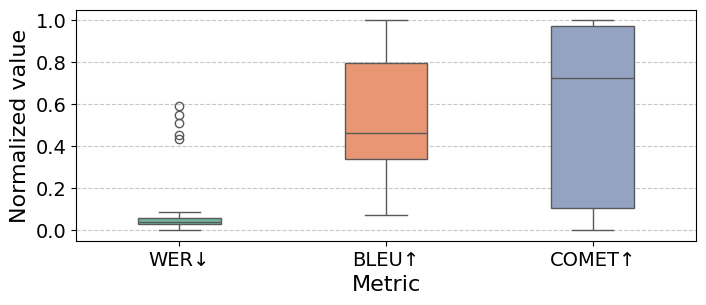}
    \caption{WER for Whisper evaluation and BLEU and COMET for LLM-translation evaluation.}
    \label{fig:metrics}
\end{figure}

Figure~\ref{fig:metrics} presents the intelligibility results, with WER evaluated for Whisper, and BLEU and COMET scores for the translation LLM. The reported WER agrees with the values found in the literature. The reference WER for this Whisper model is approximately 9.5\%, while our dataset evaluation yielded a median value of 4.5\%. This slightly improved performance may be attributed to the dataset-specific characteristics, as no modifications were made to the model.  

Regarding translation quality, the BLEU score shows a median of around 0.5, indicating that the translations are not highly reliable and contain significant errors. Meanwhile, the higher COMET score suggests that translations are generally acceptable despite existing errors, with some sentences being mistranslated. The translation direction is from English to Spanish. While Llama 3.3 does not explicitly report BLEU or COMET scores for Spanish-to-English, it scored 91.1 in the Multilingual MGSM (zero-shot) test \cite{shi_language_2022}, demonstrating strong overall performance.  

The discrepancy between our results and those reported in the literature may be primarily due to how our pipeline processes the data. Our evaluation does not merely compare two languages but assesses real-world speech records. To minimize latency, some segments may be truncated, and pauses made by the speaker may be misinterpreted as sentence boundaries. Although an additional validation LLM is used to mitigate these issues, the results remain below the state of the art. This bottleneck in translation quality has been identified as a critical area for improvement, and future iterations of the system should focus on enhancing this aspect.  

However, research has indicated that automatic translation metrics may be unreliable for evaluating translation quality \cite{zouhar-etal-2024-pitfalls}, as human evaluation remains a superior approach. Our observations confirm this: in many cases, translations are perfectly understandable, yet the metrics assign poor scores due to word order changes or structural variations introduced by the LLM. We report the standard metrics used in the recent literature without highlighting their absolute values, as we concur that their relevance is limited.

\subsection{Voice cloning subjective evaluation} 

We conducted subjective evaluations to assess the quality of the generated speech based on user perception, following \cite{MOStest}. Participants were required to use wired headphones, have no hearing impairment, and complete the test in a quiet environment. We also collected information about their native language and proficiency in Spanish: speaking, writing, and listening.

The evaluation followed a standardized MOS rating system implemented using the GoListen platform \cite{GoListen}. The online test consisted of 28 questions across two parts: (1) synthesized audio without reference to the original human recordings, \cite{text1}, and (2) comparisons between synthesized and original recordings, \cite{text2}. Each question was rated on a scale of 1 (poor) to 5 (excellent). The audio samples included phonetically balanced sentences.

\begin{figure}[b!]
    \centering
    \includegraphics[width=\linewidth]{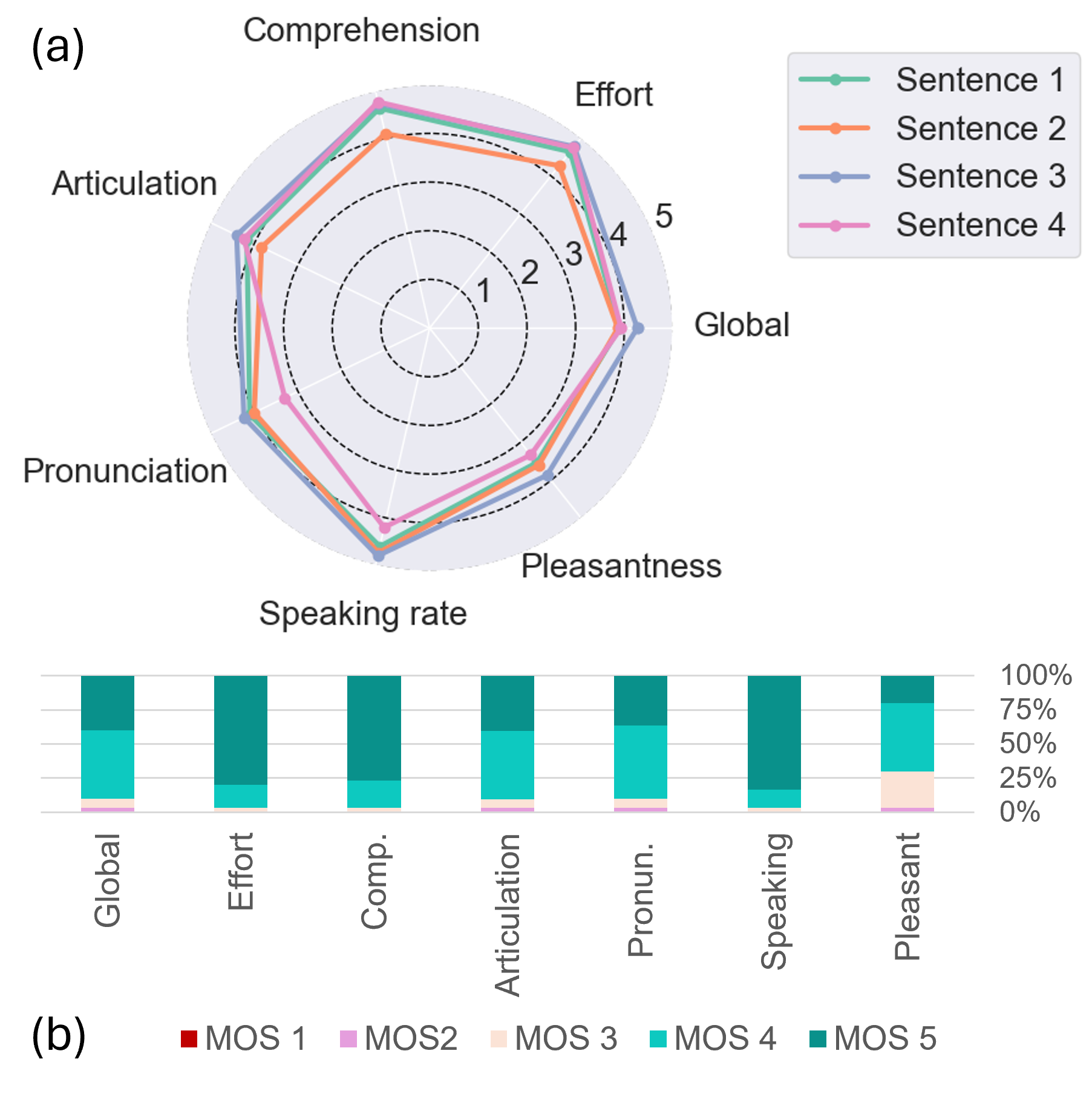}
    \caption{Results of the subjective tests: (a) chart on all sentences, (b) stacked ratings on Sentence 3.}
    \label{fig:araña}
\end{figure}

The analysis was structured into two parts. First, we evaluated items across all subjects per sentence. Second, we analyzed the dispersion for each sentence on all evaluated items. Thirty evaluators completed the test. Sentences 1 and 3 received the highest ratings overall, as shown in Fig.\ref{fig:araña}a. Sentence 3 particularly excelled in listening effort, comprehension, and speaking rate, with most ratings at 5 (Fig.\ref{fig:araña}b). However, pleasantness scored slightly lower. Sentence 4 showed good overall quality but required more listening effort and had articulation issues. Sentence 2 performed well in comprehension but struggled with pronunciation and speaking rate.

In summary, evaluators rated the synthetic audio highly, especially compared to original human recordings. These promising results highlight the potential of our solution while identifying areas for improvement, such as voice pleasantness and articulation. For further details, refer to the test\footnote{\url{https://golisten.ucd.ie/task/mushra-test/67d296bf75e8b306f88e3908}} and to the published repository.\footnote{\url{https://mateocamara.github.io/s2s}}

\begin{table*}[!htpb]
    \centering
    \caption{Measured performance parameters of the FM transmitter at various auditory locations}
    \label{tab:hardware-tests}
    \begin{tabular}{lccc}
        \hline
         Location & Peak Power (dBm) & Signal Power (dBm) & SNR (dB) \\
        \hline
        Antenna & -53.8 & -56.0 & 59.1 \\
        First Row Middle (Best Case) & -75.4 & -77.6 & 47.3 \\
        Interior (Avg $\pm$ S.D.)$^*$ & -82.0$\pm$7.3 & -83.1$\pm$7.0 & 38.2$\pm$5.7 \\
        Far Corner (Worst Case) & -93.2 & -95.1 & 33.7 \\
        Adjacent Room & -106.0 & -108.3 & 20.5 \\
        \hline
    \end{tabular}

    \vspace{0.5em}
    \small $^*$ Average room values were computed by excluding the antenna and adjacent room locations.
\end{table*}

\subsection{Hardware Tests}

To validate the performance of our FM transmitter, we assessed operational reliability and signal confinement within the designated area. We collected our measurements in the ETSIT-UPM auditorium, where irregular wall absorption, physical obstructions, and ambient noise affected propagation. The room can accommodate up to 195 people. The floor plan is 12.50~m $\times$ 14.60~m. The transmitter antenna was centered in the stage 2~m high, oriented toward the audience to optimize coverage.

In this controlled environment, we used a calibrated handheld spectrum analyzer N9342C with a test antenna from the ETS Lindgren Model 7405 kit. We used a set of reference signals, a constant 440Hz (A note), white noise, and a 20-second-long speech record. Measurements were collected in strategic locations within the auditorium to capture spatial variations in signal performance and assess leakage in adjacent rooms. 

Average measures on signal-to-noise ratio (SNR), frequency band leakage, and signal strength are listed in Table~\ref{tab:hardware-tests}. In the auditory area, the SNR was consistently above 30~dB, which indicates good reception quality. In contrast, measurements outside the auditorium produced SNR values around 20~dB, suggesting minimal signal quality; however, the observed leakage remains below acceptable interference thresholds. We also observed frequency shifts of 33.31 $\pm$ 10.2~kHz, which falls within reasonable limits. A subjective test with three volunteers provided uniformly positive feedback on audio clarity.

% \begin{table*} 
%     \centering
%     \caption{Measured performance parameters of the FM transmitter at various auditory locations}
%     \label{tab:hardware-tests}
%     \resizebox{\textwidth}{!}{%
%         \begin{tabular}{l
%                         cc  % Peak Power
%                         cc  % Signal Power
%                         cc} % SNR
%             \hline
%              & \multicolumn{2}{c}{Peak Power (dBm)} & \multicolumn{2}{c}{Signal Power (dBm)} & \multicolumn{2}{c}{SNR (dB)} \\
%             Location & No Mod. & With Mod. & No Mod. & With Mod. & No Mod. & With Mod. \\
%             \hline
%             Antenna & -40.83 & -53.75 & -41.04 & -56.01 & 74.04 & 59.09 \\
%             First Row Middle (Best Case) & -57.70 & -75.44 & -57.98 & -77.64 & 68.33 & 47.29 \\
            
%             Interior (Mean)$^*$ & -72.09$\pm$7.80 & -82.01$\pm$7.32 & -72.34$\pm$7.78 & -83.12$\pm$7.05 & 56.03$\pm$6.54 & 38.24$\pm$5.67 \\
%             Far Corner (Worst Case) & -81.24 & -93.25 & -81.48 & -95.07 & 48.36 & 33.72 \\
%             Adjacent Room & -96.68 & -105.98 & -96.95 & -108.33 & 32.60 & 20.53 \\
%             \hline
%         \end{tabular}
%     }

%     \vspace{0.5em}
%     \footnotesize $^*$ For the mean, values were computed by excluding the antenna and adjacent room locations.
% \end{table*}

\section{Conclusion} \label{sec:conclusion}

We successfully designed, implemented, and tested a system for multilingual translation and cloned speech synthesis, adapting state-of-the-art open-source components. Users can listen to our broadcast using standard radio or Bluetooth devices.

In our tests on English-to-Spanish translation, the system introduced a 5-second delay in the communication, with an average of 4.5\% WER in the transcription, 0.5 BLEU, and 0.75 COMET scores, respectively. 

From our subjective tests, we conclude that the results achieved an average rating of 4.20 on the MOS scale. In addition, our hardware tests verified reliable signal confinement and robust performance in the real environment under varying conditions, with an average SNR of 38.2 dB inside the auditorium.

The proposed system is subject to several improvements. The MeloTTS essentially conditions the latency. Optimizations to the model, particularly in how the models are handled internally, could reduce this number. The two LLM models, used for sentence completion and translation, may be integrated to accelerate execution without compromising performance. 

% For bibtex users:
\renewcommand*{\bibfont}{\small}
\printbibliography[heading=mybib]

\end{document}